# On the Effect of Magnetospheric Shielding on the Lunar Hydrogen Cycle

O. J. Tucker[1]*, W. M. Farrell[1], and A. R. Poppe[2]

[1]NASA Goddard Space Flight Center, Greenbelt, Md, USA.

[2]Space Sciences Laboratory, University of California, Berkeley, CA, USA

*Corresponding author: Orenthal Tucker (orenthal.j.tucker@nasa.gov)

NASA Goddard Space Flight Center, Magnetospheric Physics/Code 695,

Greenbelt, MD 20771, USA

**Key Points:**

- Low latitude nightside OH is depleted from waning gibbous to waxing crescent because of shielding when traversing the magnetotail.

- Low latitude dayside OH is decreased in the tail compared to out but the difference cannot be resolved with current observations.

- Magnetospheric shielding decreases the global $H_2$ exosphere by an order of magnitude during the full Moon.

**Abstract**

The global distribution of surficial hydroxyl on the Moon is hypothesized to be derived from the implantation of solar wind protons. As the Moon traverses the geomagnetic tail it is generally shielded from the solar wind, therefore the concentration of hydrogen is expected to decrease during full Moon. A Monte Carlo approach is used to model the diffusion of implanted hydrogen atoms in the regolith as they form metastable bonds with O atoms, and the subsequent degassing of $H_2$ into the exosphere. We quantify the expected change in the surface OH and the $H_2$ exosphere using averaged SW proton flux obtained from the Acceleration, Reconnection, Turbulence, and Electrodynamics of the Moon's Interaction with the Sun (ARTEMIS) measurements. At lunar local noon there is a small difference less than ~10 ppm between the surface concentrations in the tail compared to out. However, the cumulative effect of traversing the magnetotail is a surface concentration depleted by ~60 ppm for a region of the surface that spends most of its time in the lunar wake (waning gibbous to waxing crescent) on the nightside during time periods when Moon is exposed to unperturbed solar wind. The $H_2$ exosphere decreased by approximately an order of magnitude while in the magnetotail due to the decreased proton flux. The model results are consistent with preliminary exospheric observations obtained by both the (Lyman Alpha Mapping Project on the Lunar Reconnaissance Orbiter) LAMP observations and mass spectrometer measurements by the Chandrayaan - I Altitudinal Composition Explorer (CHACE).

**Plain Language Summary**

We examine how water is produced globally over the lunar surface as it orbits in/out of the magnetotail. Due to the interaction of the solar wind (SW) with Earth's magnetic field, upstream the magnetic field is compressed down to ~10 Earth radii. However, the diverted stream of SW around Earth's magnetic field results in an extended depleted region of SW protons (positively charged hydrogen) out to 1000's of Earth radii, referred to as the magnetotail. The Moon orbits at a distance of ~40 Earth radii; therefore, upstream it is within the SW, but downstream it is partially shielded while in the magnetotail during full Moon. SW protons penetrate lunar soil particles and some H atoms can chemical react with oxygen to form water-like molecules such as $OH/H_2O$. Most of the H atoms bounce around within grains until finding another hydrogen atom, chemically combine, and then escape the grain as $H_2$ into the thin atmosphere. We developed a model to calculate the global distribution of OH produced in the lunar surface and $H_2$ released to the atmosphere as the Moon orbits in/out of Earth's magnetotail. The model results are in good agreement with available observations.

**1 Introduction**

The hydrogen cycle on the Moon is a potential resource for lunar exploration. The solar wind (SW) is the dominant source of hydrogen delivered to the lunar environment, and hydrogen is primarily lost in the form of $H_2$ by thermal escape on the order of hours. Recent observations indicate that fractional amounts (10s – 100s ppm) of water are distributed over a large area of the lunar surface even at local noon [Bandfield et al. 2018; Clark 2009; Hendrix et al. 2019; Li and Milliken 2017; Pieters et al. 2009; Sunshine et al. 2009; Wöhler et al. 2017]. However, analyses of the observations cannot distinguish between whether hydrogen is present in the surface as OH or $H_2O$, and the abundance and spatial distribution of $OH/H_2O$ also remain in question [Bandfield et al. 2018; Hendrix et al. 2019; Li and Milliken 2017; Wöhler et al. 2017]. To this end, theoretical models are important tools for gaining insight about the observations and governing dynamics [Farrell et al. 2017; Jones et al. 2018; Starukhina 2006; Tucker et al. 2013]. In Tucker et al. [2019], we used a Monte Carlo model to show the diffusional lifetime of implanted hydrogen hindered by defects and forming temporary bonds with oxygen atoms is consistent the global maps of OH content extracted from IR reflectance spectra obtained by the Chandrayaan - I Moon Mineralogy Mapper ($M^3$) presented in Li and Milliken [2017]. On top of that, the model result for the content of $H_2$ subsequently degassed into the exosphere is consistent with observations by the Lyman Alpha Mapping Project (LAMP) on the Lunar Reconnaissance Orbiter (LRO) [Chin et al. 2007].

It is often considered that if the Moon's surface hydroxylation is derived from the solar wind, then the dayside surface concentration should be markedly lower during time periods within the magnetotail compared to outside of it; however, this process has not been investigated quantitatively over a full lunar orbit using a dynamic model of hydrogen implantation. Therefore, we expand our previous modeling effort to track the change in surface OH and exospheric $H_2$ as the Moon traverses the magnetosheath and magnetotail. The incident proton flux is constrained using averaged ion spectra obtained from the Acceleration, Reconnection, Turbulence, and Electrodynamics of the Moon's Interaction with the Sun (ARTEMIS) measurements [Angelopoulos 2011; Sibeck et al. 2011]. The model results show temporal hydroxyl variations at low latitudes are driven both by a thermal dropout, as a result of the diurnal temperature variation, and a magnetotail dropout, due to the decreased proton flux during traversal of the magnetotail. Thermal dropout tracks with the subsolar point for which the hydroxyl concentration is a minimum value at local noon. However, there is little difference between the hydroxyl content expected in (near full Moon) and out (near new Moon) of the magnetotail for time periods near local noon. The magnetotail dropout region is tied to selenographic coordinates as a result of the synchronous orbit about Earth. Over a lunation the concomitant effect is a localized remnant region of depleted hydrogen content that persists on the nightside from waning gibbous to waxing crescent. Potentially this region can be observed on the nightside with LIDAR at 3 μm wavelengths. Molecular hydrogen in the exosphere is expected to track with variations in the incident solar wind flux similar to helium. Model results of near surface $H_2$ during full and new Moon, respectively, are consistent with both LAMP observations and mass spectrometer measurements by the Chandrayaan - I Altitudinal Composition Explorer (CHACE) [Cook et al. 2013; Thampi et al. 2015].

**2 Background**

    2.1 Observations of Surficial OH/$H_2O$

So-called airless bodies under exposure of the SW naturally undergo ion implantation that in turn alters both physical and chemical properties of the surface. On the Moon, implanted SW protons are of particular interest because the neutralized H atoms can combine with oxygen producing water products (OH/$H_2O$) that could be a potential resource in aid of human exploration. Near Infrared (NIR) spectra of reflected sunlight from the Moon's surface obtained by three different missions possess a spectral absorption feature around ~2.8 – 3.0 μm providing evidence of widespread water products over the lunar surface [Bandfield et al. 2018; Clark 2009; Grumpe et al. 2019; McCord et al. 2011; Pieters et al. 2009; Sunshine et al. 2009]. However, each analysis of NIR spectra uniquely depends on the estimate of the thermal emission component which must be removed in order to resolve the reflectance spectra, and the extracted OH/$H_2O$ spatial coverage differ significantly between analyses [Bandfield et al. 2018; Li and Milliken 2017; Wöhler et al. 2017]. Bandfield et al. [2018] reported that the absorption feature was present at all latitudes and independent of the time of day. They concluded that time of day and latitudinal dependences could not be explicitly extracted from the spectra, whereas Wöhler et al. [2017] produced global maps showing a mild time of day dependence in the absorption band depth at mid to high latitudes with the minimum occurring near local noon. In contrast to the Wöhler et al. interpretation, Li and Milliken [2017] determined that the spectra possessed a significant time of day dependence varying up to ~200 ppm at mid-latitudes. Their orbit averaged spectra also revealed a strong latitudinal dependence with surface concentrations as low as 0 ppm at low latitudes, at latitudes above ~ 70° the concentration increased to ~ 500 – 700 ppm.

Sparse observational data of surface hydration during times when the Moon was within Earth's magnetotail have been obtained by LAMP and $M^3$. Likewise, using the same thermal model as Li and Milliken [2017], Cho et al. [2018] reported that $M^3$ spectra obtained in the tail are indicative of surface concentrations that were only slightly less than that when out of the tail at latitudes < 10 degrees. However, they note that there is not enough data to conclusively demonstrate the presence of a shielding effect. They identified 21 data strips with similar properties observed during local noon for which 3 were obtained within the magnetotail and 19 obtained out of the magnetotail. The absorption feature while in the magnetotail was ~ 15 % weaker than the observations out of the magnetotail. Likewise, Li et al. [2018] also found a small decrease in the absorption feature in the tail but at mid latitudes. Spectral slopes extracted from reflected FUV (164 – 173 nm) spectra of the surface by LAMP have also provided evidence of dayside surface hydration [Hendrix et al. 2019]. Hendrix et al. reported that the LAMP data are consistent with a diurnal variation of in surface density of ~$10^{16}$ m$^{-2}$ (< 1% of a monolayer) two hours about lunar local noon at latitudes of 30 – 50°. However, their spectra in and out of Earth's magnetotail were similar and they concluded that the OH/$H_2O$ surface concentration is not solar wind induced. It is important to note that LAMP approximately samples regolith depths on the order of 10 – 100 nm [Hendrix et al. 2019], whereas $M^3$ samples down the regolith down to millimeter depths [Li and Milliken 2017]. In addition, Hendrix et al. noted that LAMP obtained spectra of surface regions periodically both in and out of the magnetotail so the effects of shielding may have been averaged out.

2.2 Observations of OH, $H_2O$ and $H_2$ in the Exosphere

Observations of water products in the lunar exosphere are an additional constraint to surface hydration models. The Apollo Lunar Surface Experiments Package ALSEP of Apollo 17 included a mass spectrometer, referred to as the Lunar Atmospheric Composition Experiment (LACE), which recorded spectra with small peaks for mass 17 and 18 amu molecules [Hoffman 1975, Figure 17-3) in the Taurus Littrow Valley (TLV); however, the measurements were deemed as artifacts produced by instrument outgassing [Hoffman 1975]. Only an arbitrary upper limit of $H_2O$ density of $2 \times 10^7$ cm$^{-3}$ is given in the Apollo 17 preliminary science report. Furthermore, the report does not discuss how this upper limit was obtained. LACE observations only conclusively identified the noble gases He, Ne, and Ar according to the final mission report [Hoffman 1975]. Regarding additional water products such as $H_2$ and $O_2$ they were also contaminated by instrument outgassing, but upper limits of < 65000 cm$^{-3}$ and < 200 cm$^{-3}$ were derived respectively as discussed in Hoffman et al. [1975].

Cook et al. [2013] derived upper limits of atmospheric gases using a mission averaged UV spectrum obtained by the Lyman Alpha Mapping Project (LAMP) on the Lunar Reconnaissance Orbiter (LRO) for a collection of so-called twilight observations (two hours before dawn and two hours after dusk). LAMP observed reflected UV light in the wavelength range of 1050 - 1650 Å to infer averaged $H_2$ surface densities. The observed brightness was converted to surface densities using exosphere theory assuming a surface temperature of 120 K (e.g., Chamberlain 1963; Cook et al. 2013 equations 1 - 6). Because of exospheric global transport driven by the diurnal surface temperature, LAMP observations cannot be directly compared with LACE measurements that occurred on the nightside at 20º latitude in TLV. Nevertheless, *qualitatively* the LAMP observations of the abundance of H (24 cm$^{-3}$), O (5.4 cm$^{-3}$) and $H_2$ (1000 ± 500 and 1400 ± 500 cm$^{-3}$) [Cook et al. 2013; Stern et al. 2013] are comparable with the LACE measurements and upper limit inferences.

The CHACE mass spectrometer sampled the lunar atmosphere during full Moon (November 14, 2008), on the dayside at altitudes of ~ < 94 km for lunar latitudes of > 20º S [Sridharan et al. 2010]. OMNI data of the solar wind indicated typical solar wind and interplanetary magnetic field conditions for most of the day, so the observation likely occurred while the Moon was inside of the Earth's geomagnetic tail. CHACE measured the noble gases (Ne and Ar) at levels consistent with LACE, however the measurements of He and $H_2$ (500 – 800 cm$^{-3}$) were depleted in comparison to observations by LAMP and LACE, likely because the Moon was in the magnetotail. In addition, the near surface dayside exospheric densities for $H_2O$ were found to be on the order of ~$10^6$ cm$^{-3}$ [Sridharan et al. 2010, 2015]. However, the effect of outgassing from the spacecraft on the measurements during the mission was not well characterized during its operational phase.

NASA's Lunar Atmosphere and Dust Environment Explorer (LADEE) sampled the Moon's exosphere at equatorial region at latitudes within ~ 20º of the Moon's equator for 7 months in 2013 – 2014. When accounting for the location and time of day they also measured consistent values of the noble gases compared to LACE, LAMP and CHACE [Benna et al. 2015]. LADEE measured a near surface density of water vapor of 0.62 cm$^{-3}$ [Benna et al. 2019]. We note these densities are much lower than measured by CHACE. To date, LADEE measurements of $H_2$ in the exosphere have not been analyzed in detail.

2.2 Surface Hydration Models

Several theoretical investigations have examined dynamics of the lunar environment leading to the production hydrogenated molecules in the lunar surface and their subsequent loss mainly focusing on the formation and diffusion of H, $H_2$, OH and $H_2O$ [Farrell et al. 2017; Farrell, Hurley, and Zimmerman 2015; Grumpe et al. 2019; Jones et al. 2018; Starukhina 2006]. Starukhina [2006] predicted that solar wind implanted elements would accumulate in the polar regions due to the lower degassing rates occurring at lower temperatures by using Fick's laws of diffusion. In this formulation, diffusing H atoms are hindered by a variety of chemical and physical traps produced in the top layers of the regolith due to constant irradiation. Then, any H atoms that diffuse to the surface will have some given probability to chemically combine as $H_2$ and degas into the exosphere. Chemical trapping occurs when hydrogen forms chemical bonds with other atoms bound to silicate grains. Due to the abundance of oxygen within lunar soil, it is expected that many diffusing H atoms will attach to an unsatisfied valence electron (dangling bond) of a bound O atom forming metastable OH. On the other hand, physical trapping refers to the mobility of H atoms being hindered by crystal defects such as interstitials and vacancies. Chemical trapping typically has a higher activation energy barrier to diffusion than physical trapping. Starukhina [2006] highlighted that extracting a diffusivity prefactor, $D_0$, to characterize the rate of hydrogen diffusion in lunar silica minerals under constant irradiation is challenging because the variety of trapping sites are expected to have a broad range of activation energies, $E_a$. For example, in one scenario if the chemical traps in a regolith particle are saturated, diffusion would proceed relatively quickly because the only barrier to diffusion would be physical traps. However, in another scenario a regolith particle may have recently freed chemical trapping sites due to irradiation, and diffusion would proceed slower. The diffusivity is defined by the following Arrhenius' relationship, $D = D_0\exp(-E_a/kT)$, where $k$ is Boltzmann's constant and $T$ is the surface temperature. Published results of the diffusion rates of hydrogenated products in low temperature lunar soils under irradiation are not currently available but such programs are underway, e.g., McClain et al. [2020].

Farrell et al. [2015; 2017] investigated the abundance of hydrogen in the lunar soil by using a distribution of activation energies constrained to experimental results of hydrogen degassing rates from radiation damaged amorphous silica [Fink et al. 1995]. They demonstrated the latitudinal and diurnal trends observed the IR spectra of the surface [Clark 2009; McCord et al. 2011] were consistent with this approach when using the Fink et al. [1995] diffusivity prefactor of $D_0 = 10^{-12}$ m$^2$/s and a Gaussian distribution of activation energies characterized by a mean energy and width of $F(E_a = 0.5$ eV, $E_w = 0.1$ eV). Tucker et al. [2019] adapted this approach for use in a global Monte Carlo model that tracked the surface accumulation of H and exospheric degassing of $H_2$ over several lunations. They obtained a quasi-steady state surface concentration of H *quantitatively* consistent with the latitudinal and diurnal trends extracted from the M$^3$ IR spectra shown in the global maps by Li and Milliken [2017], and the subsequent result of degassed $H_2$ in the exosphere was consistent with LAMP observations [Cook et al. 2013]. In that study, it was assumed that hydroxyl formulation is primarily limited by diffusion, and therefore, the hydrogen densities were directly correlated with the IR signature. As an example, they increased the activation energy to $U = 0.7$ eV to slow diffusion, and found far more retention of H in the surface, a loss of any diurnal effect, and a far lower $H_2$ exospheric density than measured by LAMP.

Jones et al. [2019] examined the reaction kinetics leading to the production and destruction of hydroxyl that ultimately result in the release of water into the exosphere via recombinative desorption by numerically solving a coupled set of chemical kinetic rate equations. In this approach, it is assumed that implantation results in the production of detached hydroxyl defects that recombine with other surface bound hydroxyls and degas as $H_2O$. The resulting surface concentrations are consistent with a buildup of hydroxyl at high latitudes, but the abundances are much smaller than the Li and Milliken [2017] M$^3$ distributions. Recombinative desorption is active on the lunar dayside at temperatures of ~300 – 400 K and highly efficient at temperatures > ~ 600 K [Jones et al. 2018]. Micro-meteoroid impacts can also provide the thermal energy required for conversion [Zhu et al., 2019].

Grumpe et al. [2019] adapted the H atom diffusion model of Farrell et al. [2017] to include OH formation, OH diffusion and OH photolysis. Using the diffusion model, they fit the M$^3$ maps presented in Wöhler et al. [2017] which inferred hydroxyl content on order of 1 - 10 ppm with mild diurnal and latitudinal variations. They concluded this absorption signature is best fit the using an average activation energy of 0.4 - 0.45 eV for H diffusion and an OH photolysis lifetime of $5 \times 10^4 – 1.5 \times 10^5$ s.

Each of described models points to solar wind implantation as the source of the widespread 3-micron feature. However, there remains disagreement on the dominant mechanism responsible for the production of OH/$H_2O$. As discussed above, observed degassed products in the exosphere can provide an additional constraint. Both the Grumpe et al. and Jones et al.

models produce a temporal dayside OH/H$_2$O exosphere in excess of the upper limits of water vapor in the lunar exosphere of < ~0.62 cm$^{-3}$ inferred from LADEE [Benna et al. 2019]. That is, in the absence of meteoroid impacts LADEE derived an upper limit for a permanent steady state water exosphere. To this end, models of hydrogenated surface products should be considered in parallel with the degassed hydrogen products to the exosphere. However, to date there have only been two published measurements of water in the lunar exosphere.

## 3 Methods

In our previous study, we provide a more detailed overview of the diffusional and exospheric model that is used in this study [Tucker et al. 2019]. The model tracks the flux of solar wind/ magnetosphere protons implanted into the surface, and the subsequent hydrogen diffusional lifetime and loss as H$_2$ in the Moon's exosphere as a function surface temperature. We adapted this model to take advantage of the Moon's synchronous rotation in order to roughly account for the change in the flux of protons as the Moon passes through the magnetosheath and terrestrial magnetotail as observed by ARTEMIS. To this end, the model is carried out in a Moon-centered frame where longitudes 0° and 180° define the peak full Moon and New Moon phase, respectively. Therefore, the Moon is defined to be in the magnetotail when the subsolar longitude rotates from 30° - 0° and 360° - 332°, and the dawn and dusk magnetosheath flanks are defined to be within longitudes 57° - 30° and 332° - 304°, respectively (see e.g., Figure 3, Poppe et al., [2018]). All other time periods are assumed to be out of the magnetosphere with the Moon exposed to typical solar wind conditions.

When the Moon traverses Earth's magnetosphere it is exposed to shocked solar wind plasma within the magnetosheath, and ambient plasma both of solar origin and ion outflow produced from Earth's atmosphere within the magnetotail [Poppe, Farrell, and Halekas 2018]. Since 2011, the twin ARTEMIS spacecraft have collected data on the electromagnetic environment near the Moon during elliptical orbits with perigees ranging from 10 – 1,000 km above the surface and perigee of ~10-12 lunar radii [Angelopoulos, 2011; Sibeck et al., 2011]. Poppe et al. [2018] averaged ARTEMIS ion energy spectra along the lunar orbit around the Earth for over 5 years in geocentric solar ecliptic coordinates. Outside of the Earth's magnetospheric boundaries the Moon is bombarded by a mean solar wind flux of 1.9 × 10$^{12}$ m$^{-2}$ s$^{-1}$. While inside the magnetosphere, the data indicate that the Moon spent an average time of ~2.1 days in the dusk and dawn magnetosheaths combined exposed to a shocked mean proton flux of 2.4 × 10$^{12}$ m$^{-2}$ s$^{-1}$, and 4.8 days in the magnetotail exposed to a decreased, but non-zero flux of 2.2 × 10$^{11}$ m$^{-2}$ s$^{-1}$. We estimated the time spent within magnetospheric boundaries using the Poppe et al. [2018] (their *Figure 3*) taking the average time of the lunar synodic period (Full Moon to Full Moon) of 29.5 days.

The individual diffusive pathway of each implanted H atom is characterized by a lifetime obtained by using a Gaussian distribution of activation energies to account for the variety of defects expected in regolith particles [Farrell et al. 2017, 2015; Tucker et al. 2019]. Based on the range of energies of *incident* protons derived from ARTEMIS measurements (Poppe et al. [2018], *their figure 4*) we calculated the implantation depth using results from the Stopping Range of Ions in Matter (SRIM) Monte Carlo program [Ziegler, Ziegler, and Biersack 2010]. SRIM was used to track incident protons within an SiO$_2$ amorphous target, and the average implantation depth ranged between ~5 – 100 nm with an averaged depth of ~20 nm for incident energies of ~100 eV – 10 keV. The diffusional lifetime $\tau = h^2\exp[E_a/kT]/D_0$ depends on the implantation depth *h*, activation energy $E_a$, local surface temperature *T* and diffusional coefficient prefactor, $D_0$. Tucker et al. [2019] found that the $M^3$ global distributions of surficial OH [Li and Milliken 2017] could be reproduced by using diffusional lifetimes characterized by Gaussian energy distribution function with a peak activation energy of $E_a = 0.5$ eV and width of $E_w = 0.08$ eV, and a diffusional prefactor of $D_0 \sim 10^{-12}$ m$^2$ s$^{-1}$ [Fink et al. 1995]. Therefore, we used these model parameters for all simulations carried out in this work.

At lunar temperatures hydrogen most readily desorbs in a molecular form, e.g., $H_2$, because hydrogen is very reactive in atomic form [Starukhina 2006]. We assumed that implanted hydrogen predominantly desorbs as $H_2$, which is consistent with exospheric observations by LACE [Hodges, 1973], LAMP [Cook et al. 2013; Stern et al. 2013] and CHACE [Thampi et al. 2015]. Once in the exosphere, representative $H_2$ particles are tracked in the Moon's gravitational field assuming full thermal accommodation to the local surface temperature and no adsorption. The representative particles have a statistical weight derived from the solar wind flux $F(Z) = n_{sw}v_{sw}\cos(Z)$, where $n_{sw}$ and $v_{sw}$ are the solar wind density and velocity, respectively and $Z$ is the solar zenith angle. The desorbed hydrogen molecules are prescribed velocity vectors using the Maxwell-Boltzmann flux speed distribution and cosine angular distribution [Brinkmann 1970]. If they have energy to escape the gravitational potential or have probability to be loss by photoionization or photodissociation, referred to here as photodestruction, they are removed from the simulation. We adopted a $H_2$ photodestruction lifetime of ~50 days from Huebner and Mukherjee [2015]. The loss is dominated by thermal escape on the order of hours once degassed from the surface. Below, we present results on how loss by escape and variability of the source rate over a lunation as the Moon passes in/out the geomagnetic tail affect the hydroxyl production and exospheric content.

## 4 Results

The simulations were carried out for 15 – 20 lunations in which we obtained surface maps of the quasi-steady state surface concentration of OH and steady state exospheric abundances. At latitudes below ~ 70°, the OH content is balanced with H implantation and $H_2$ desorption, however at high latitudes particles with activation energies in the tail of the distribution >> 0.5 eV can accumulate over geological time scales. In reality, there are a finite amount of high activation energy trapping sites (dangling O bonds), so once those sites become saturated the distribution will effectively shift to lower energy physical trapping sites. The saturation oxygen density for the lunar regolith has been estimated to be $\sim 10^{16} – 10^{17}$ cm$^{-2}$ [Schaible and Baragiola 2014; Starukhina 2006]. After 20 lunations, our simulations resulted in a surface density ~3 orders of magnitude less than the saturation limit, with only ~$10^{14}$ cm$^{-2}$ at the highest latitudes > ~ 80 degrees. Saturation in the simulation would occur in ~$10^3$ years, however the accumulation would also be limited by sputtering and meteoritic impacts. The concomitant effect of space weathering on the hydrogen desorption rate will be considered in more detail in a future study.

Figure 1 shows a schematic of the solar illumination of the Moon's surface in a sun-fixed frame compared to the magnetotail dropout region, which is fixed in selenographic coordinates over the lunar orbit. The magnetotail dropout region (MDR) is defined in selenographic coordinates as the region of the surface that is illuminated during peak full Moon within the magnetotail and therefore receives less proton flux, labeled as 'A' in Figure 1. The location of this region is shown over the lunar orbit in Figure 1 to identify its relative position to the thermal dropout region. The thermal dropout region (TDR) refers to the illuminated region of the surface. As expected, during full Moon the combined effect of the lower solar wind flux and thermal dropout results in a comparatively decreased dayside surface concentration of OH. However, not so intuitive is the presence of a 'ghost feature' within the MDR that never gets a full dose of the solar wind because when the Moon exits the tail it is aligned obliquely to the SW flow and for most of the orbit this region is within the lunar wake and not directly exposed to the solar wind.

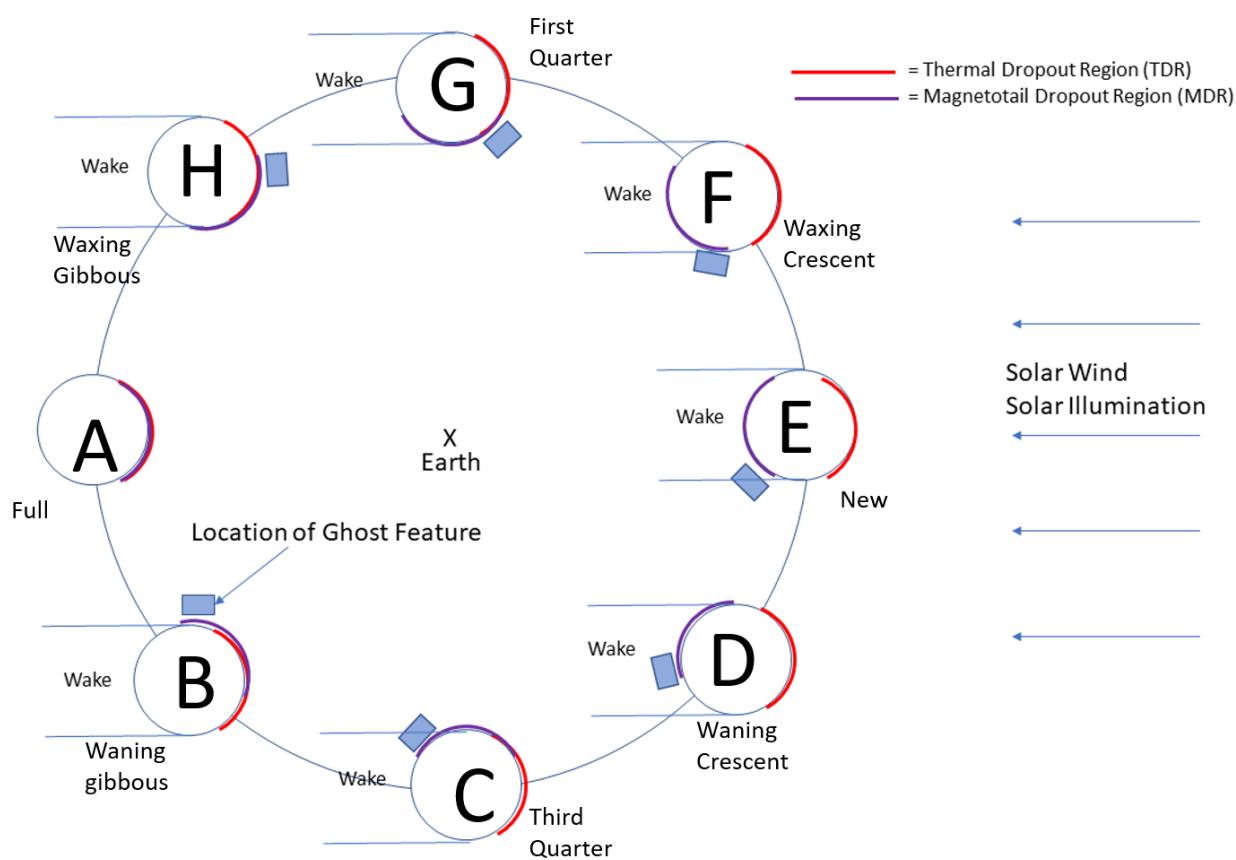

**Figure 1**. Schematic of the thermal (TDR) and magnetotail (MDR) dropout regions shown for each of the lunar phases. The blue box identifies a region of the surface within the MDR that collectively receives the least amount of proton flux over the lunar orbit, and remains a surface OH depleted region over a lunation. The sun is located to the right.

Snapshots of the hydroxyl concentration for the 8 lunar phases are shown in Figure 2. As previously described in Tucker et al. [2019], the maps show both an accumulation of hydroxyl with increasing latitude up to <~ 1000 ppm, and a time of day dependence varying up to 100 - 200 ppm, consistent with Li and Milliken [2017]. During the full Moon phase the dayside surface concentrations are the lowest, Figure 2A, and during the waxing gibbous the dayside surface concentrations are largest, Figure 2H. The ghost feature is seen in Figure 2B. During the waning gibbous phase it appears on the lunar nightside (i.e., the low density lunar wake). It persists undisturbed on the nightside (wake) until new Moon (Figure 2E) for our assumption of a negligible nightside source of protons. Once the Moon enters the waxing crescent phase the ghost feature is once again on the dayside (TDR) and in the unperturbed solar wind, thus now rotating from the nightside into the direct solar wind. The model clearly indicates that a surface remnant of depleted hydrogen from the magnetotail passage remains 'in memory' on the surface throughout most of a lunation. It remains primarily in the wake region on the lunar nightside, and thus does not get hydrogenated until it rotates back into the solar wind flow (near F and G).

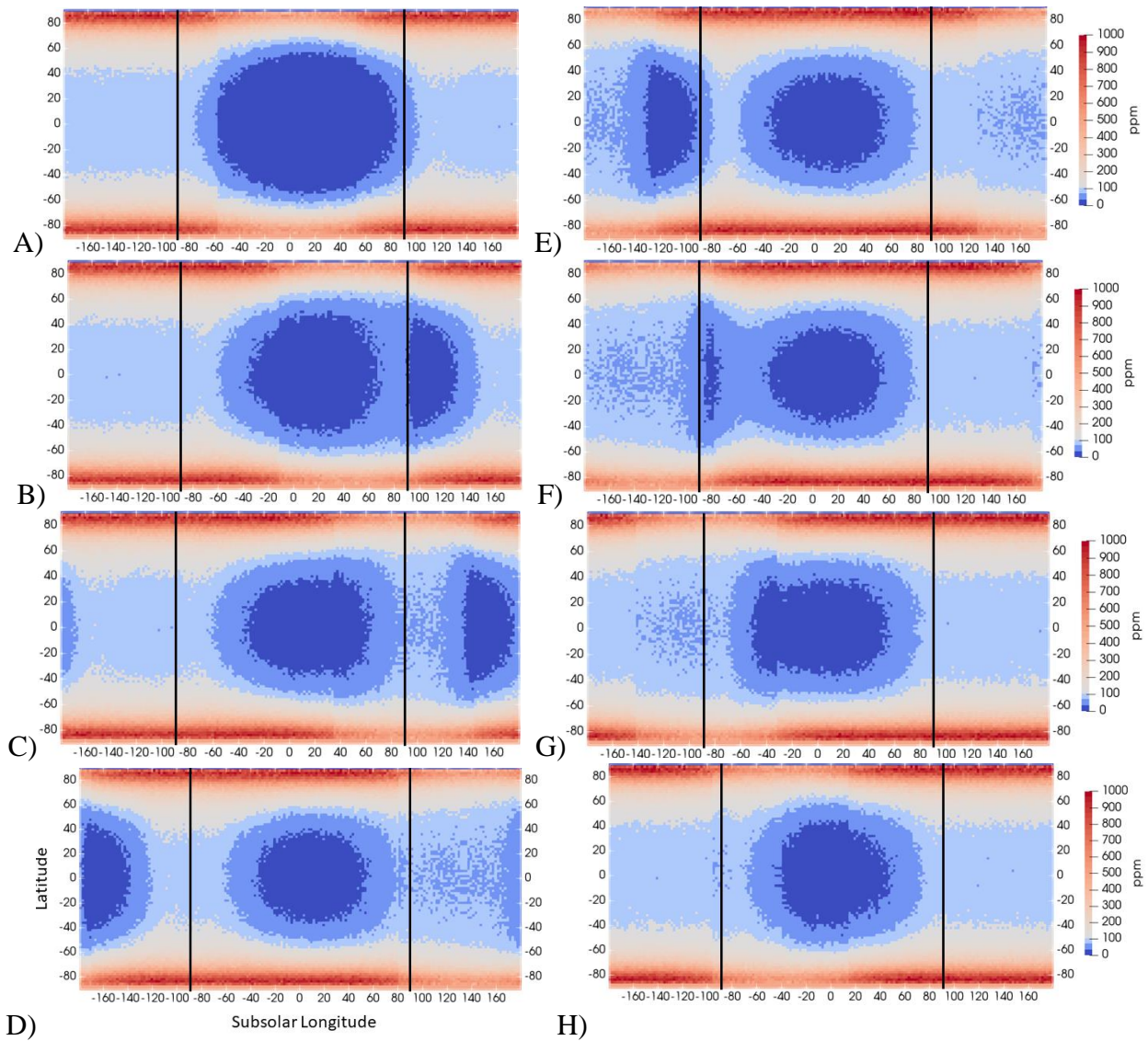

**Figure 2**. Diurnal maps of modeled OH surface density (left) as a function of subsolar longitude and latitude. The subsolar point is located at 0 degrees longitude, and the morning and evening terminators are located at -90 and 90 degrees longitude, respectively. A) – H) represents the 8 lunar phases shown in Figure 1, where A) is full Moon and E) is new Moon.

There are few observations of the IR absorption feature for the same coverage areas in and out of the geomagnetic tail. Examining $M^3$ data strips, Cho et al. [2018] identified 3 strips in the magnetotail and 19 strips out of the magnetotail observed by $M^3$ with similar maturity, mineralogy and location that were also observed at a similar time of day. They found the strength of the absorption feature in the tail compared to out of the tail was weaker by ~10% at local noon. However, this difference is within the uncertainties of both sets of observations (in tail and out of tail), so the findings were inconclusive. Our simulations estimated OH surface concentrations of ~5 ppm compared to ~18 ppm at local noon for the full Moon and new Moon phases, respectively. Even though the model results indicate a 70% difference in the respective concentrations such differences in these fractional amounts are likely not discernable. That is, the reported detection limit of $M^3$ was 20 ppm in the Li and Milliken [2017] analyses. It is important to note that the concentrations in the Grumpe et al. [2019] analyses are more than 2 orders of magnitude less, highlighting the significant disparity in the interpretation of $M^3$ spectra. Nevertheless, the model results suggest that the effect of shielding on the OH surface concentration is not discernable by comparing the dayside distributions when in and out of the tail. There is a lower *nightside* concentration of OH on the near side compared to the *nightside* concentration of the far side during full Moon. The predawn nightside surface concentration has a minimum value of ~ 12 ppm at low latitudes during new Moon (Figure 2E) compared to ~ 80 ppm during full Moon (Figure 2A). However, because these regions are un-illuminated there are no observational data for comparison.

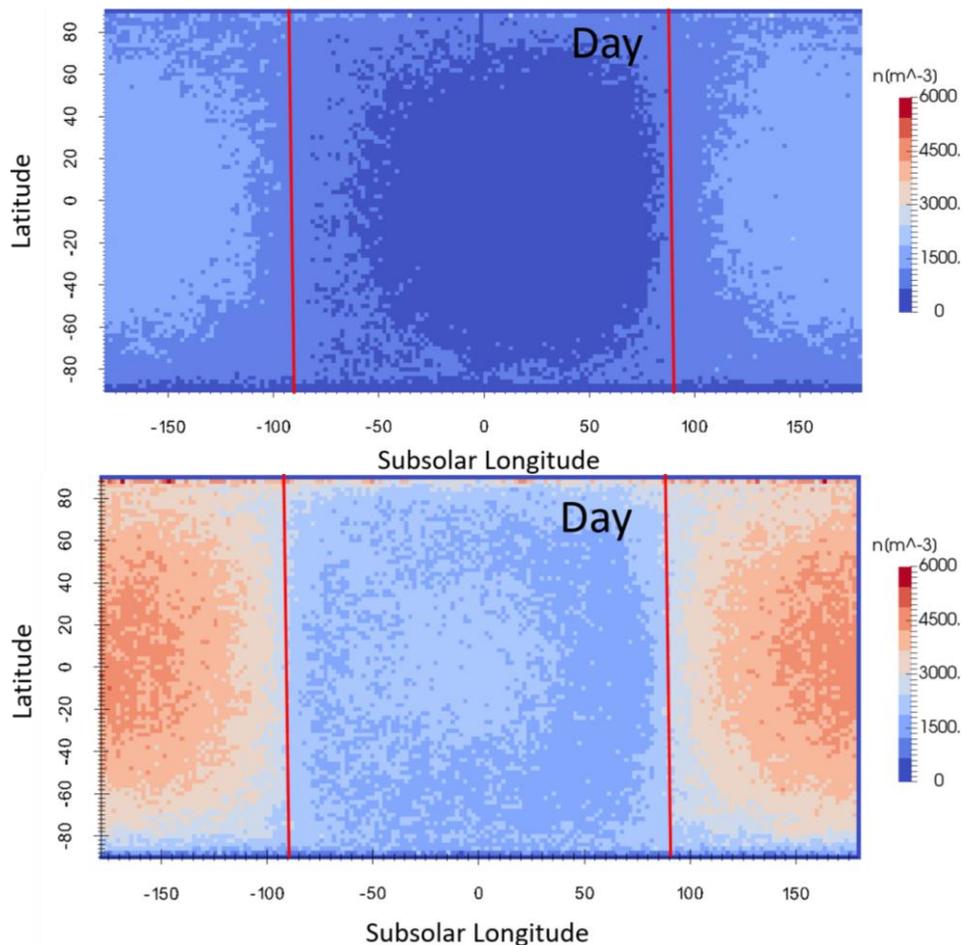

**Figure 3**. Diurnal maps of the near surface exospheric $H_2$ density as a function of subsolar longitude and latitude for full Moon/tail (top) and new Moon/solar wind (bottom) The subsolar point is located at 0 degrees longitude, and the morning and evening terminators are located at -90 and 90 degrees longitude, respectively.

The effect of shielding on the $H_2$ exosphere is much more significant due the short timescale for thermal escape, which is on average of the order of hours. Figure 3 shows an order of magnitude lower global $H_2$ exospheric densities when in the magnetotail. However, once the Moon is out of the magnetospheric tail and once again exposed to the magnetosheath and unperturbed SW the exospheric densities increase within hours due to the high degassing rates.

We calculated a globally average mass loss rate of 0.026 kg/s of molecular hydrogen from the Moon which is comparable to the source rates of solar protons ~ 0.032 kg/s, ~ 0.038 kg/s and ~ 0.0035 kg/s in the nominal solar wind, magnetosheath, and magnetotail, respectively. This result is consistent with the loss rate of 0.002 – 0.028 kg/s derived in Hurley et al. [2017] for the conversion rate of solar wind proton implantation to $H_2$ in mineralogical complex lunar soil obtained from LAMP observations. For completeness, we note a recent laboratory simulation derived a lower loss rate for $D_2$ of ~ 0.0009 – 0.0013 kg/s obtained from an experimental analog of deuterium implantation into olivine [Crandall, Gillis-Davis, and Kaiser 2019].

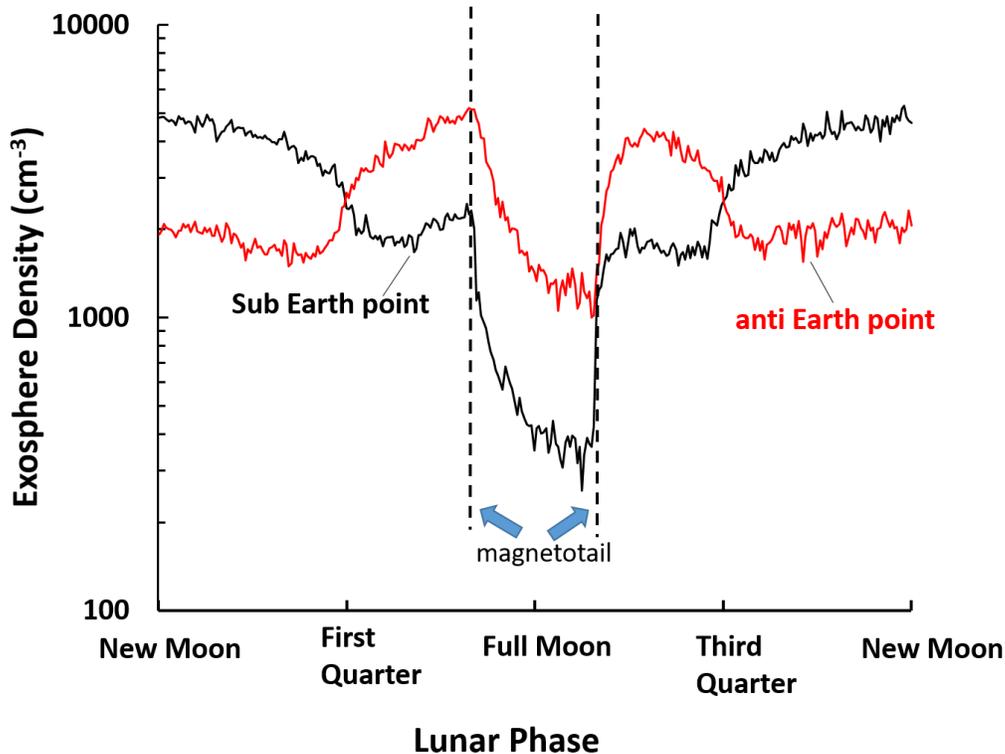

**Figure 4**. Near surface exosphere density as a function of the lunar phase at the sub Earth point (black curve) and the anti-sub Earth point (red curve).

Both LAMP and LADEE observations show that lunar helium (mass 4 amu) in the exosphere is directly correlated to the solar wind because the densities have been observed to decrease proportionally to the change in alpha particle flux proton ($He^{++}$) flux when the Moon traverses the geomagnetic tail [e.g., Benna et al. 2015]. We show a similar response of the $H_2$ near surface exosphere densities at the sub Earth and anti-sub Earth points in Figure 4. As expected, the time series of $H_2$ exosphere densities at the sub/anti-sub Earth points differ because in the tail the sub Earth point is on the lunar dayside and the anti-sub Earth point is on the night side. Due to the surface temperature, on the dayside $H_2$ has a scale height a factor of ~4 larger on the dayside compared to the nightside.

These results are consistent with preliminary observations of exospheric $H_2$ by LAMP presented in Cook et al. [2016]. Their analyses appear to show a general decrease in $H_2$ densities when the Moon was in the tail, however the effect was deemed inconclusive [Cook et al. 2016]. For the spectral range sampled by LAMP (575 – 1965 Å) the $H_2$ spectrum is much weaker than He and the variations in the observations are comparable to the inferred densities [Cook et al. 2016]. LAMP twilight observations of He clearly show a depletion of exospheric densities of ~ 8000 cm$^{-3}$ compared to 16000 cm$^{-3}$ in and out of the tail, respectively. However, the averaged LAMP H$_2$ observations and their corresponding variance are on the same order of ~1000 cm$^{-3}$. Figure 4 shows that the decrease of H$_2$ in the tail is on the order several 1000 cm$^{-3}$, so the current observations cannot conclusive identify a decrease in density in the tail. More data are required to reduce the spread in the observations, which might be obtained with the extended LRO mission. We predict H$_2$ densities near the terminators on the order of ~500 and ~1500 - 2000 cm$^{-3}$ in and out of the tail, respectively. These results indicate that a measurement precision of < ~100 cm$^{-3}$ are required to clearly resolve the change in H$_2$ density at the terminators in and out of the geomagnetic tail.

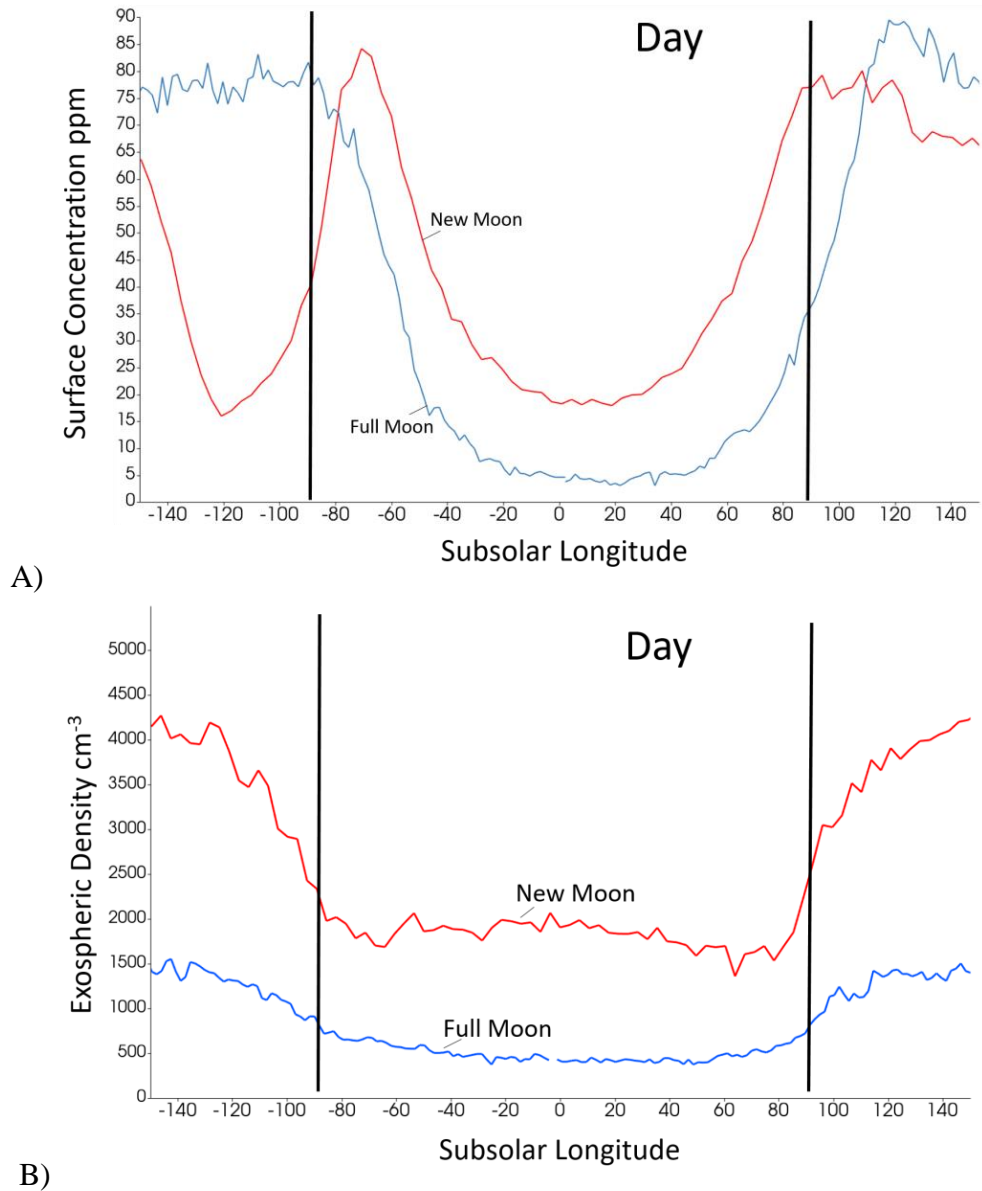

A)

B)

**Figure 5**. Time of day equatorial profiles for OH surface concentration (left) and H$_2$ exospheric density (right) during new Moon (red curve) and full Moon (blue curve). The terminators are identified by the black vertical lines (-180 degrees is the morning terminator).

In summary, using ARTEMIS data of the mean proton flux in the magnetospheric sheath $2.4 \times 10^{12}$ m$^2$ s$^{-1}$ and tail environments $2.2 \times 10^{11}$ m$^{-2}$ s$^{-1}$ we calculated that at the subsolar point the surface OH concentration is only lower by ~10 ppm from when in the tail compared to out (Figure 5A). Likewise, the near-surface exospheric density of H$_2$ decreases by an order of magnitude at the subsolar point from ~2000 cm$^{-3}$ to ~ < 400 cm$^{-3}$ during the ~5 day time period spent in the tail (Figure 5B). Without a replenishing source of protons, H atoms in low energy trapping sites, characterized by diffusional activation energies that are < ~0.5eV, are lost during the time period spent in the tail. That is, due to the short diffusive and thermal escape lifetimes of ~1 hour (loss rate of ~$7.8 \times 10^{24}$ H$_2$/s), it is expected that both the surface concentration of OH and global distribution of H$_2$ in the exosphere respond relatively quickly with changes in the solar wind flux.

## 5 Conclusions

A major challenge of understanding the lunar hydrogen cycle is the range of differing interpretations of M$^3$ spectra of the 3 μm signature on the lunar surface. Li and Milliken [2017] inferred surface concentrations up to ~0 - 1000 ppm that increased with latitude and exhibited diurnal variations up to 200 ppm at mid to low latitudes, whereas Wöhler et al. [2017] derived concentrations on order < ~10 ppm with little diurnal variation and variation with increasing latitude. Bandfield et al. [2018] reported that local variations in the absorption feature were modest and near the limit of uncertainty of the correction to the extraction method applied. To this end, theoretical models based on the physics governing the implantation process can provide insight on the interpretation of the observations [Grumpe et al. 2019; Jones et al. 2018; Tucker et al. 2019].

Our results are consistent with the global interpretation of the M$^3$ spectra as presented in Li and Milliken [2017]. M$^3$ has limited observations of similar surface regions at similar times comparing the OH content in and out of the tail [Cho et al. 2018]. Recently, Hendrix et al. [2019] presented more robust comparisons of LAMP observations in and out of the tail of FUV reflectance spectra that also characterized surface hydration. They found that during the lunar day there was relatively little difference between the spectra, and thus concluded that the diurnal hydration signature is due to migration and not prompt solar wind implantation.

Here, we have shown for solar wind implantation when the thermal dropout region and magnetotail dropout region are aligned there is a reduction in surface hydroxyl. However, the relative difference between the hydroxyl content on the dayside in and out of the tail is small and not readily discernable using current IR observations. From the full Moon to waxing crescent phases, there is a 'ghost' feature on the nightside for which the surface concentration is an almost an order of magnitude lower than the nightside densities during waxing crescent until full Moon. This feature is a remnant of a portion of the lunar surface that never receives a full dose of unperturbed SW flux because during most of the time that the Moon is immersed in the solar wind, this region is in the lunar wake on the nightside. Near first Quarter, this region rotates to the dayside, but before rotating back to the nightside it merges again with the thermal dropout region. There are no observations of the shallow surface (< ~100 nm) nightside surface concentrations.

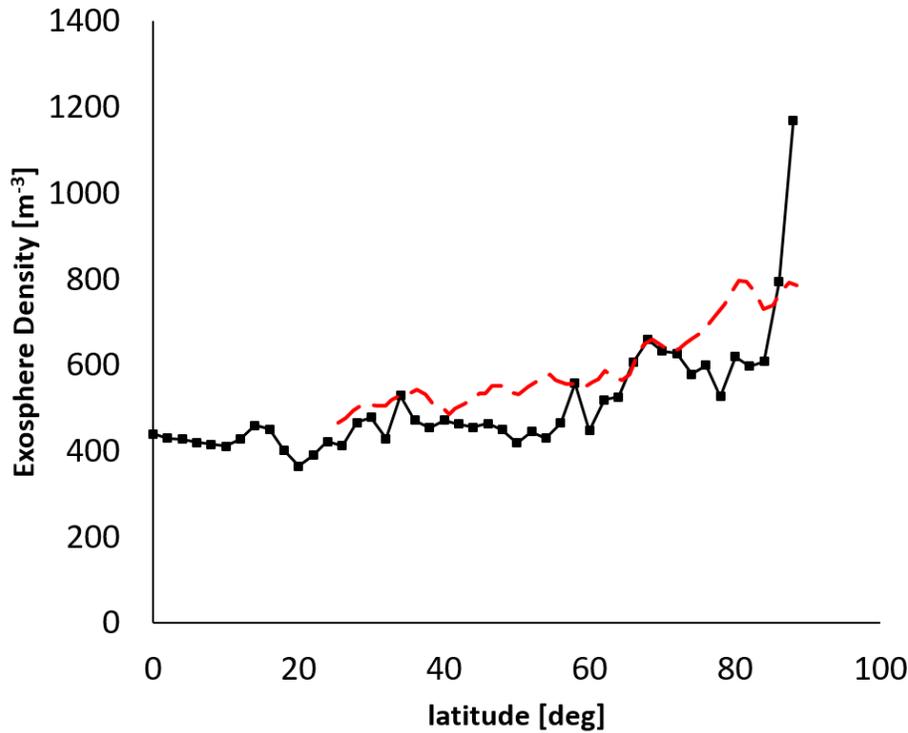

**Figure 6**. Near surface exosphere density versus latitude. The dashed red curve shows CHACE $H_2$ data digitized from Thampi et al. [2015] extrapolated to the surface from in situ measurements of the exosphere while the Moon was in the magnetotail. The corresponding model data shown by the solid black curve with squares is consistent with the CHACE observation.

There are little data of $H_2$ densities in the lunar exosphere for when the Moon is in the magnetotail. Our results are consistent with mean LAMP data of the exospheric densities inferred near the terminator, however these LAMP analyses are ongoing [Cook et al. 2016]. Nevertheless, the modeled distribution of $H_2$ is consistent with the CHACE observation of $H_2$ while in the magnetotail. CHACE mapped dayside exospheric densities at altitudes below ~90 km over latitudes 20 – 80° to the surface, shown in Figure 6. Note that Thampi et al. [2015] attributes the smaller fluctuations seen in Figure 6 to local surface topography on the kilometer scale and not to variations in local surface temperature. Any small-scale correlations between the model and the CHACE data are likely fortuitous considering the large hopping distance of $H_2$ molecules at lunar temperatures, and the altitudes at which CHACE sampled the exosphere. The Monte Carlo model does not include topography effects so any physical fluctuations would be related to the local surface flux, surface temperature, and mean hopping distance. More simulations are required to determine whether the density variations are physical or just due to statistical fluctuations of the Monte Carlo approach. New observations to be obtained by the Chandrayaan-2 orbiter CHACE2 mass spectrometer may provide new insight on the $H_2$ exospheric distribution over the full lunar cycle.

Overall, solar wind implantation remains the likely driver of the lunar hydrogen cycle. As shown in Tucker et al. [2019] and again here, an important constraint to consider in chemical and/or diffusional models of hydrogen products with the lunar surface is the abundance of degassed products within the exosphere. Both the Jones et al. and Grumpe et al. methods require the degassing of a significant amount of OH or $H_2O$ on the lunar dayside, although they did not directly estimate the resulting amounts. Similarly, Hendrix et al. [2019] predict a dayside density of $H_2O$ of ~500 $cm^{-3}$ if ballistic transport is responsible for the diurnal variation observed by LAMP of the absorption feature. To date, there has not been a significant detection of $OH/H_2O$ in the lunar exosphere. LADEE has placed the sole constraint on mass 17/18 species in the exosphere on the order of < ~0.62 $cm^{-3}$ [Benna et al. 2019], far below that postulated by Hendrix et al. [2019]. However, our modeling shows that a high volume end-state for the implanted solar wind is in the form of outgassing $H_2$, with surface emission of $H_2$ from a solar wind implantation and Fink-like diffusion process consistent with both the character of the surface OH (as analyzed by Li and Milliken, 2017] and $H_2$ exosphere [Cook et al. 2013, 2016; Thampi et al. 2015].

We note that an additional method of validation of this model would be a detection of the ghost feature of the magnetotail depletion that remains on the nightside during the late phase/waning gibbous periods. There are ongoing developments to use a 3 micron LIDAR to examine polar craters and such a LIDAR should be capable of detecting this nightside OH depleted ghost feature. If found, it would add further evidence for the solar wind- H diffusion- OH formation- $H_2$ exosphere interconnection.

**Acknowledgments and Data**

The model data used to create all plots in this publication are available on the 4TU.Centre for Research Data repository, doi:10.4121/uuid:578ee2d8-9811-4bb9-826e-553232dd7271. This work was supported by SSERVI DREAM2 and LEADER. The numerical simulations were carried out using the XSEDE high performance computers. The authors gratefully acknowledge Dr. Larissa Starukhina useful email exchanges that contributed to this work. A.R.P. acknowledges support from the NASA ARTEMIS mission under grant #NAS5-02099. O.J.T. dedicates this work to Richard A. Tucker Sr. (1933 - 2020) for his enduring encouragement.